\documentclass [twocolumn, aps, showpacs] {revtex4}
\usepackage{graphicx}
\input{epsf}
\begin{document}
\draft
\title {Spin-nematic order in the frustrated pyrochlore-lattice quantum rotor model}
\author {Karol Gregor, David A. Huse and S. L. Sondhi}
\address{Department of Physics, Princeton
University, Princeton, NJ 08544}
\date{\today}
\begin{abstract}
As an example of ordering due to quantum fluctuations, we examine
the nearest-neighbor antiferromagnetic quantum $O(n)$ rotor model on
the pyrochlore lattice.  Classically, this system remains disordered
even at zero temperature; we find that adding quantum fluctuations
induces an ordered phase that survives to positive temperature, and
we determine how its phase diagram scales with the coupling constant
and the number of spin components.  We demonstrate, using quantum
Monte Carlo simulations, that this phase has long-range spin-nematic
order, and that the phase transition into it appears to be first
order.
\end{abstract}
 \maketitle

\section{Introduction}
Ordering of antiferromagnetic spins on geometrically frustrated lattices
is a subtle problem \cite{moessner-rev,ramirez-rev}.
By definition, the canonical nearest neighbor antiferromagnetic
interaction on such lattices fails to produce a unique classical
ground state. Thus low temperature ordering in such systems must
be attributed to additional interactions or to selection via thermal
or quantum fluctuations, the latter phenomenon being termed order
by disorder.

\begin{figure}[h]
\epsfxsize=3in \centerline{\epsfbox{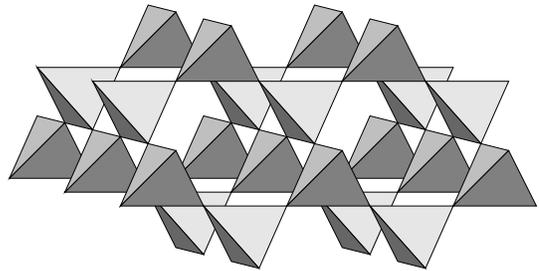}}
\caption{Pyrochlore Lattice} \label{fig_pyronew}
\end{figure}

In recent years there has been much interest in antiferromagnetism
on the pyrochlore lattice, Figure \ref{fig_pyronew}
\cite{oleg-website}. The classical statistical mechanics of the
purely nearest neighbor problem is now fairly well understood. It is
known \cite{moessnerchalker} that XY spins order, by disorder,
collinearly \cite{fn-xy} while Ising, Heisenberg and
higher-dimensional spins do not order even in the zero temperature
limit. The correlations of this set of cooperative paramagnets have
been found to exhibit a universal dipolar form characteristic of an
underlying gauge field \cite{isakov1}, truncated by a correlation
length that diverges as $T \rightarrow 0$.

While thermal fluctuations thus do not lead to ordering in the
Heisenberg problem, there is much work arguing that quantum
fluctuations do lead to ordering at low temperatures, for the case
of Heisenberg spins. Close to the classical limit, this conclusion
follows from arguments based on the $1/S$ expansion discussed
recently by Henley \cite{Henley}. Henley derives an effective
Hamiltonian on the space of classical ground states which captures
the effects of the zero point energy of harmonic spin waves in a
loop expansion. This indicates a selection of collinear ground
states with a large unit cell and a residual degeneracy that is of
order $O(L)$, where $L$ is the linear size of the sample. It is
expected that a nonlinear treatment of the spinwaves will lift this
remaining degeneracy and predict long range spin order in a
particular collinear configuration. In the opposite limit of small
spins and large quantum fluctations, there is a set of
investigations principally of the $S=1/2$ case, starting with the
pioneering work of Harris \textit{et al.}
\cite{Harris,Tsunetsugu,Tsunetsugu2,KogaKawakami,BergAltmanAuerbach}
which suggest a breaking of the inversion and translation symmetries
of the lattice with only short range order among the spins. For
$S=1$ Yamashita {\it et al.}\cite{YamashitaUedaSigrist} have also
argued for a breaking of inversion symmetry as well as long range
order in the transverse component of the spin chirality. Two large
$n$ studies \cite{Sondhi,msg} have also suggested symmetry breaking
for small spin values. Indeed Ref.~\cite{msg} finds that the quantum
dimer model that arises in a large $N$ treatment, does indeed break
inversion symmetry and further translational symmetry breaking via
an order by disorder mechanism.

In this paper we further explore the impact of quantum fluctuations
on spins on the pyrochlore lattice by endowing them with the
dynamics of quantum rotors instead. The resulting problems, which
are readily defined for $O(n)$ symmetric spins for all $n$, are
distinct from the Heisenberg spin problems even at $n=3$ although
there is clearly a family resemblance which detailed analysis will
bear out. The rotor models are interesting in their own right
\cite{Sachdev} and their $n=1$ representative is the transverse
field Ising model which has been studied on a variety of frustrated
lattices \cite{MoessnerSondhi}.

The Hamiltonian for the quantum rotor model on the pyrochlore
lattice is given by
\begin{equation}
H = g\sum_i \vec{L}_i^2 + \sum_{<i,j>} \vec{S}_i \cdot \vec{S}_j~,
\label{eq_H}
\end{equation}
where $\vec{S}_i$ is a unit vector in $n$ dimensions located at
site $i$, $L_i$ is its associated angular momentum and the second
sum goes over all nearest-neighbor pairs of sites on the
pyrochlore lattice. The coupling constant $g$ measures the strength
of quantum fluctuations.

At any $n$ and in the limit of large coupling constant $g$, the
ground state has all rotors in their zero angular momentum state,
and thus no long range correlations or symmetry breaking. In this
limit there is also a gap of $O(g)$ so this quantum paramagnetic
phase is stable for a range of values of $g$. \footnote{The robust
existence of the quantum paramagnetic phase makes rotors quite
different from spins where large quantum fluctuations do not select
a unique state. One can, however, consider modified rotor models
where monopoles placed at the centers of rotors modify their low
energy spectra. Such a rotor model is, for example, the lattice
representation of the $O(3)$ non-linear sigma model with a
topological term in $d=1+1$ \cite{shankar-read}.} The opposite limit
of $g=0$ is the classical model, which has a highly degenerate
ground state and, for $n>2$, dipolar correlations in the
zero-temperature limit but no significant strength in any Fourier
component of the magnetization \cite{moessnerchalker,isakov1}.

\begin{figure}[h]
\epsfxsize=4in \centerline{\epsfbox{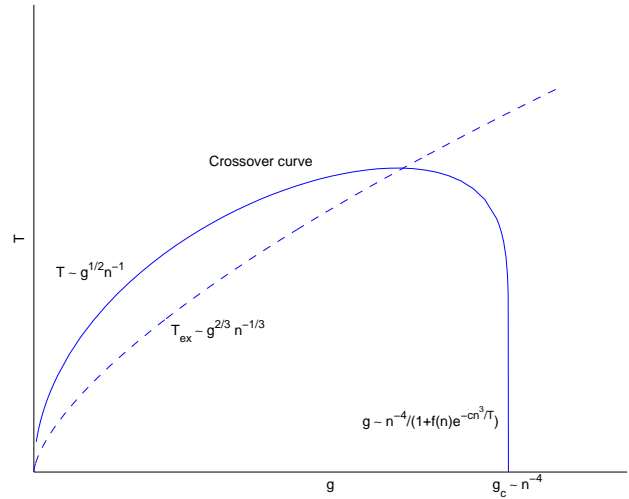}}
\caption{Schematic crossover diagram of a single tetrahedron, with
the scalings with $g$, $T$ and $n$ indicated.  The solid line is
where the system crosses over from being collinear to disordered,
while the dashed line shows the value of $g$ where, for a given $T$,
the deviations from collinearity are minimized. The solid line is
also a very approximate phase transition curve for the whole system.
However we expect that the curve starting at $g_c$ will bend
initially to the right due to the entropy of the spin waves that
exist in the ordered phase.} \label{fig_crossover_diagram}
\end{figure}

The problem at hand is to understand how the system interpolates
between these two very different disordered limits. In the following
we report progress on this question. Primarily we will show that the
system develops spin-nematic (collinear) long-range order for a
finite range of values of $g \in (0,g_c)$ at $T=0$. More generally
we derive the schematic phase diagram in the temperature-coupling
constant plane shown in Fig.~\ref{fig_crossover_diagram} with the
various scalings that we have derived indicated. We have not been
able to establish or exclude further symmetry breaking at the lowest
temperatures into a state with long range spin order. We do however
offer analytic evidence and simulational evidence (for the case
$n=3$) that such further symmetry breaking is likely quite weak.

We turn next to understanding the quantum mechanics of the building
block of the pyrochlore lattice, the single tetrahedron. The basic
results derived here will next enable us to deduce the existence of
nematic order on the full lattice and the scalings of various energy
scales with $n$. Thereafter we describe results of a simulation
for the $n=3$ problem and conclude with a discussion of the $1/n$
expansion for this problem and a summary.

\section{Single Tetrahedron}

In this section we analyze the basic unit of the pyrochlore lattice
--- a single tetrahedron. It consists of four mutually coupled spins
and its hamiltonian is
\begin{equation}
H = g\sum_{i=1}^{4} \vec{L}_i^2 + \sum_{1 \le i < j \le
4}\vec{S}_i\cdot\vec{S}_j~.
\end{equation}
This may also be viewed as a system of four interacting particles,
each moving on the surface of a sphere in $n$ dimensions. The first
term in $H$ is their kinetic energy, while the second term is a
repulsive interaction potential.

In the classical limit of $g=0$ it is known \cite{moessnerchalker}
that for $n=2$ (XY spins) the spins order collinearly in the zero
$T$ limit --- two spins pointing in one direction and the other two
in the opposite direction \footnote{The probability distribution
becomes more and more centered at collinear situation as $T \to 0$.
Of course there is no spontaneous symmetry breaking in this finite
system.}. On the other hand, for $n>2$ the spins remain disordered
even in the $T\rightarrow 0$ limit as we review below. In the rest
of this section we show that in the quantum case ($g>0$) for any $n$
there are ways to take the $g\rightarrow 0$ and $T\rightarrow 0$
limits that give collinearly ordered spins. Furthermore, we find the
dependence of the crossover from ordered to disordered spin states
as a function of $g$, $T$ and $n$, as indicated in Figure
\ref{fig_crossover_diagram}.

\subsection{Classical spins, $g=0$}

\begin{figure}[h]
\epsfxsize=2.5in \centerline{\epsfbox{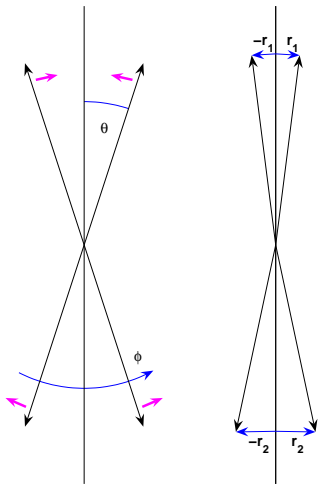}} \caption{a)
Parametrization of the ground state, showing the four spins and the
reference axis.  All the spins are at angle $\theta$ from the
reference axis.  The short arrows indicate the mode that goes soft
as the spins become collinear. 
The line with no arrows is the reference axis in
spin space that we measure $\theta$ from.} \label{fig_vectors1}
\end{figure}

We start by describing the configurations of the spins and the
classical ground states (cgs). The spins are in a cgs if the
potential is minimized; this is when the four spins add up to
zero. This implies that in a cgs, the fourth spin lies in the
three-dimensional space spanned by the other three spins.  Thus
the ground states can be parameterized by two angles $\theta$ and
$\phi$ as shown in Figure \ref{fig_vectors1}, plus an overall
rotation in spin space. $\theta$ gives the deviation of all spins
from collinearity, while $\phi$ is the angle between the planes in
spin space spanned by each pair of spins.  All four spins in a cgs
are at the same angle $\theta$ from the reference axis; the
reference axis that minimizes $\theta$ is used in our convention
of parameterizing these ground states. Mostly we will consider
nearly collinear states with small $\theta$.

For both the classical and the quantum analysis, we need to consider
the matrix of second derivatives of the potential energy with
respect to the orientations of the four spins at a given classical
ground state.  First, this matrix generically has $(3n-4)$
eigenmodes with zero second derivative; these directions span the
local manifold of classical ground states.  For small $\theta$ it
also has $(n-1)$ eigenmodes with second derivative of order unity,
and one ``soft mode'' with second derivative proportional to
$\theta^2$. This last mode corresponds to the spin deviation away
from the nearly collinear cgs that decreases $\theta$ for one pair
of nearly parallel spins and increases it for the other pair, as
indicated in Fig. 2.  This soft mode's stiffness is independent of
the other angle $\phi$.

At low temperature, the classical system at equilibrium can be
approximated as occupying the states with potential energy within
$k_BT$ of the ground state and not those at higher potential. Due to
the soft mode, the number of such states near an almost collinear
cgs is proportional to $T/\theta$ (for $T<\theta^4$).  For $n=2$,
this concentration of the accessible states near $\theta=0$, due to
the one mode that softens there, causes the ``order by disorder''
effect and the spins order collinearly in the $T\rightarrow 0$
limit.  But the number of cgs with a given $\theta$ is proportional
to $\theta^{2(n-2)}$, due to the freedom of rotations about the
reference axis.  This means that the probability density of $\theta$
behaves as $\sim\theta^{(2n-5)}$ at small $\theta$. Thus for $n>2$
the collinear states do not dominate even in the zero temperature
limit, and the classical spins remain disordered.

\begin{figure}[h]
\epsfxsize=4in \centerline{\epsfbox{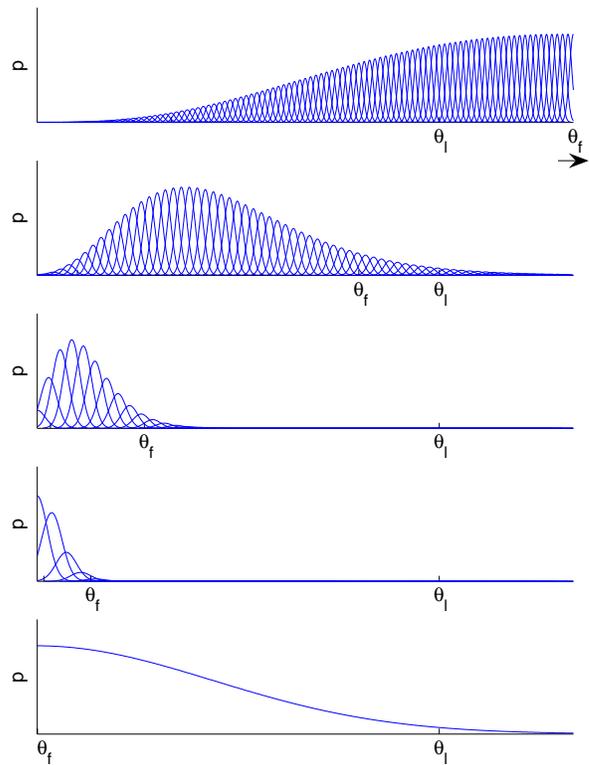}} \caption{Very
schematic drawing of evolution of the distribution of the
eigenstates and probability of the system at a fixed $T$ as $g$ is
increased. The height of each peak corresponds to the probability
that the system is at the given $\theta$. The width of each peak
corresponds to the spread of each eigenstate. The distribution is
localized approximately up to an angle $\theta_f$ and we say that
the system is localized if $\theta_f < \theta_l \sim1$. At the top
$g$ is small, the system is noncollinear and effectively classical,
occupying many excited states.  As $g$ is increased (moving down in
the figure) the system becomes more collinear, and the number of
different excited states occupied at equilibrium decreases.
Meanwhile, the spread of $\theta$ within the ground state is always
increasing as $g$ increases.  At $g_{\rm min}$ the excitation energy
of the lowest excited state passes through $T$; this is where the
system is most collinear.  As $g$ increases further, the spread
within the ground state increases (bottom).} \label{fig_wave_c}
\end{figure}

\subsection{Quantum ground state, $g>0$}
Here we show that for small $g$ the ground state wavefunction is
localized around the collinear state and we obtain the scaling with
$g$ and $n$ of its spread, its energy and the energy of the lowest
excited states.  First we present a ``power-counting'' variational
argument

We treat the quantum zero-point energy of the motion normal to the
manifold of classical ground states as that of harmonic oscillators.
For small $\theta$, there are $(n-1)$ ``stiff'' oscillators, each
with zero point energy $\sim\sqrt{g}$, and the one soft mode, with
zero point energy $\sim\theta\sqrt{g}$.  The latter produces a
$\theta$-dependent effective potential that lifts the degeneracy
within the set of classical ground states and has a minimum at
$\theta=0$. We are interested in characterizing the ground state and
the excited states in this potential.

The set of classical ground states for a given reference axis
constitutes a $(2n-3)$-dimensional manifold that intersects with
itself at the collinear states where $\theta =0$.  The angle
$\theta$ is proportional to the distance from this intersection.
Each pair of nearly parallel spins may be rotated about the
reference axis. These latter rotations give $2(n-2)$ dimensions of
motion. One can also rotate the reference axis, which gives the
remaining $(n-1)$ dimensions of motion.  In the ground state, these
reference-axis degrees of freedom are in the zero total angular
momentum eigenstate and we will generally ignore them in this
section of the paper.

Thus we consider a variational wavefunction that is localized near
$\theta=0$, spreading in angle by $\gamma$ in each direction on the
$(2n-3)$-dimensional manifolds near the collinear state.  The
kinetic energy of such a state is $\sim g(2n-3)/\gamma^2$.  The
typical angle $\theta$ in this state is the (Pythagorean) sum of
$(n-1)$ angles of displacement away from the reference axis that are
mutually perpendicular to each other and each of order $\gamma$, so
$\theta\sim\gamma\sqrt{n-1}$. At the level of accuracy we are using
now, $n\sim (n-1)\sim (2n-3)$, so the kinetic energy is $\sim
n^2g/\theta^2$, and the full variational energy is this plus the
effective potential of $\sim\theta\sqrt{g}$. Minimizing this with
respect to $\theta$ gives the following estimate for its typical
value in the ground state:
\begin{equation}
\theta_0 \sim n^{\frac{2}{3}}g^{\frac{1}{6}}~.
\end{equation}
Thus we see that the ground state is collinear ($\theta_0\rightarrow
0$) in the $g\rightarrow 0$ limit.  The crossover from collinear
ordering (small $\theta_0$) to a strongly disordered state can be
defined as occurring at $\theta_0\sim 1$, which puts the crossover
in the ground state at
\begin{equation}
g_c\sim n^{-4}
\end{equation}
for large $n$.

The contributions to the ground state energy from the motion within
the manifold of classical ground states as well as that from motion
along the ``soft mode'' direction are
\begin{equation}
E'_0 \sim (ng)^{\frac{2}{3}}~.
\end{equation}
In the former case, this energy is due to the motion in $(2n-3)$
directions, so the energy of motion in just one of these directions
along the manifolds of cgs, and thus the energy to excite the motion
in that direction to a higher-energy eigenstate is
\begin{equation}
E_{ex} \sim g^{\frac{2}{3}}/n^{\frac{1}{3}}~.
\end{equation}
There are other, lower-lying excited states that do not alter the
degree of collinearity: these involve ``rigid body'' rotations of
the ground state we are discussing, and, at even lower energy for
small $g$, tunneling between the three distinct ways of pairing the
four spins.

The above arguments have approximated the soft mode as harmonic,
which is correct for the bulk of the ground state wavefunction in
the limit of large $n$.  However, we should check that the
$g$-dependence that we have derived remains valid at small $g$ even
when $n$ is small (such as for the interesting case of $n=3$).  To
do better at small $g$ and small $\theta$ but general $n$, we
combine the soft mode degree of freedom with the manifold of
classical ground states, and formulate the Schrodinger equation
within this space.  Label the two pairs of nearly parallel spins 1
and 2 and let the displacement of one of the spins in each pair away
from the reference axis be $\vec r_1$ and $\vec r_2$, respectively.
The other spin in each pair has precisely the opposite displacement.
These displacements are $(n-1)$-dimensional vectors and, when small,
their magnitudes are equal to the angles of rotation of the unit
vectors.  To leading order at small $g$ and small $r_i$ the
resulting Schrodinger equation is
\begin{equation}
H'\psi=-2g(\nabla_1^2+\nabla_2^2)\psi+(1/2)(r_1^2 - r_2^2)^2 \psi =
E'\psi
\end{equation}
By comparing the kinetic and potential energy terms in this
Hamiltonian $H'$, we see that indeed the characteristic scale of
angle is $g^{\frac{1}{6}}$ and the scale of energy is
$g^{\frac{2}{3}}$, as obtained above in the harmonic approximation.
The ground state in this unusual quartic potential for large $n$ is
concentrated away from the collinear state at $r_1\cong r_2\sim
n^{\frac{2}{3}}g^{\frac{1}{6}}$, where the harmonic approximation
used before remains valid.  But for small $n$ the ground state
wavefunction has considerable weight close to collinearity where the
harmonic approximation is not appropriate.  The leading correction
at small $g$ to (7) appears to be from the $\theta$-dependence of
the stiffness of the other ``hard'' modes that were ignored.  This
gives a $\sim\theta^2\sqrt{g}$ contribution to the effective
potential that thus contributes to the energy at order
$g^{\frac{5}{6}}$, one order higher in our ``small'' parameter of
$g^{\frac{1}{6}}$.

\subsection{T$>$0}
Next we consider nonzero temperature for small $g$, examining the
crossovers that occur, first from the fully quantum regime for
$T<E_{ex}$ where the system remains in its nearly-collinear ground
state, to an intermediate regime ($E_{ex}<T<T_c(g)\sim
g^{\frac{1}{2}}/(n-2)$) where there are many thermal excitations
present but it remains near collinear, and then, for $n>2$, to the
disordered regime at higher $T$.

The modes within the manifold of classical ground states are excited
when the temperature reaches the excitation energy
$T_{ex}=E_{ex}\sim g^{\frac{2}{3}}/n^{\frac{1}{3}}$.  At higher
temperatures, we can treat these degrees of freedom as classical.
The ``soft mode'' has an excitation energy of $\sim\theta\sqrt{g}$
and this mode will be in its classical regime where it is highly
excited only at angles where this is less than $T$.  There the
probability of being near a particular cgs is
\begin{equation}
P(\theta)\sim\frac{T}{\theta\sqrt{g}}~;
\end{equation}
this applies for $\theta_0<\theta<T/\sqrt{g}$.  For
$\theta<\theta_0$, within the support of the ground state
wavefunction, $P(\theta)\sim P(\theta_0)$.  At larger angles,
greater than both $\theta_0$ and $T/\sqrt{g}$, the soft mode is in
its ground state, which gives an effective energy of
$\sim\theta\sqrt{g}$ and
\begin{equation}
P(\theta)\sim\exp{(-\theta\sqrt{g}/T)}~.
\end{equation}

Again, the number of distinct cgs with a given $\theta$ behaves as
$\sim\theta^{2(n-2)}$ at small $\theta$, so the typical value of
$\theta$ is near the maximum of $\theta^{2(n-2)}P(\theta)$.  In
the intermediate regime we are discussing, which is
$E_{ex}<T<T_c(g)\sim g^{\frac{1}{2}}/(n-2)$, this maximum occurs
near
\begin{equation}
\theta(T)\sim\frac{(n-2)T}{\sqrt{g}}~.
\end{equation}
Thus, for $n>2$ and small $T$ and $g$, the crossover to the
thermally disordered state occurs at
\begin{equation}
T_c(g)\sim g^{\frac{1}{2}}/(n-2)~,
\end{equation}
where $\theta(T)\sim 1$.  These small $g$, $T$ results apply as long
as the quantum ground state is itself nearly collinear, which
requires $g<n^{-4}$ and thus $T<n^{-3}$.  The crossover temperature
$T_c(g)$ must have a maximum of order $n^{-3}$ and then decrease to
zero at $g_c$, as indicated in Fig. 1.

The striking result here is that this simple system of four rotors
has a nonmonotonic, or re-entrant behavior at low $T$ as one
increases $g$ from the classical limit of $g=0$ to the quantum
limit of large $g$ for $n>2$.  This is illustrated in Fig. 3.
Initially at small $g$ it is disordered due to the large entropy
of the disordered states relative to the collinear states.  As $g$
is increased, the effective potential due to the soft mode
increases, causing the system to be more and more confined to the
nearly collinear eigenstates, as $\theta(T)$ decreases with
increasing $g$. This trend continues until the energy $E_{ex}$ of
the excited states within the manifold of cgs increases to of
order $T$, at which point the system is predominantly in the
ground state, with $\theta_0\sim T^{\frac{1}{4}}n^{\frac{3}{4}}$.
This point, where the deviations from collinearity are minimized,
occurs at $g_{\rm min}\sim T^{\frac{3}{2}}n^{\frac{1}{2}}$. Beyond
this point, further increase of $g$ (decrease of the ``mass'')
causes the ground state to instead deviate more from collinearity
with increasing $g$, until it crosses over in to the fully
disordered quantum regime at $g_c$.

\section{Pyrochlore Lattice} Now that we have examined the behavior
of a single tetrahedron, we turn to the question of the behavior of
our model on the pyrochlore lattice.  This lattice consists of a
three-dimensional array of corner-sharing tetrahedra, so adjacent
tetrahedra share a single site.  There are two aspects of the
ordering that occurs in the single tetrahedron in the
appropriately-taken small $g$, $T$ limit, namely the axis along
which the spins are all collinearly aligned, and which pairs of
spins are pointing which way along that axis.  If two adjacent
tetrahedra sharing a single spin both order, they must order along
the same axis, so this ``spin-nematic'' order should propagate
throughout the lattice. As we show below, we have found good
evidence from quantum Monte Carlo simulations that the pyrochlore
lattice model has a phase with long-range spin-nematic order in a
region of its phase diagram with nonzero $g$ and $T$.  Thus the
crossovers we discussed in the previous section for the single
tetrahedron become true phase transitions on the full lattice. Since
this is an isotropic-to-nematic phase transition with a cubic
invariant in its Landau theory, this phase transition is expected to
be first order, and that expectation is indeed supported by our
simulations.

The other aspect of the ordering does not strongly propagate
between adjacent tetrahedra:  The spin nematic order picks a
particular axis in spin space, but there are 6 different
low-energy configurations of the spins' directions along this axis
for each tetrahedron.  If a given tetrahedron orders in to one of
these spin patterns, the adjacent tetrahedra that each share a
spin with it still each have 3 spin patterns that remain
compatible with the first tetrahedron.  At this level of
consideration, for the full lattice the entropy of this spin
degeneracy is extensive, and the spins remain disordered, although
collinear. It seems likely that this spin degeneracy is lifted to
some degree by tunneling between the various spin configurations,
but this is something that so far we have not detected, either
analytically or in our simulations.

\subsection{Degrees of freedom and modes}
We start by reviewing the counting of the number of degrees of
freedom and of constraints\cite{moessnerchalker}. Let $N$ be the
number of tetrahedra. Then there are $2N$ spins and $2N(n-1)$
degrees of freedom. Each tetrahedron has zero total spin in a
classical ground state; this gives $Nn$ constraints.
Thus the dimension of the set of cgs is $N(n-2)$, and hence there
are $N(n-2)$ zero modes around a generic cgs configuration.
However, around a fully collinear configuration there are $N(n-1)$
zero modes. Thus there are $N$ soft modes whose stiffnesses vanish
as we approach the fully collinear configuration.

For a small displacement away from the collinear configuration let
$\theta$ be the average of all $\theta$'s of all spins. We will now
argue that for the soft modes the associated second derivative of
the potential goes as $\lambda \sim \theta^2$, just as for the
single tetrahedron.

\begin{figure}[h]
\epsfxsize=3in \centerline{\epsfbox{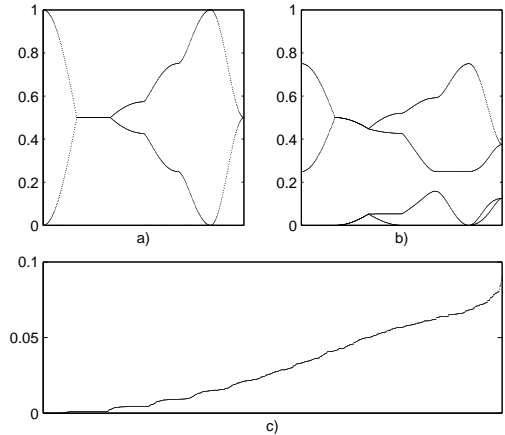}} \caption{The
eigenvalues along chosen directions for a) $\theta=0$ b)
$\theta=0.3$ and c) all eigenvalues of the two lowest bands ordered
by their magnitude} \label{fig_Bands_all}
\end{figure}

In a collinear configuration consider two zero directions
parameterized by $z_\alpha$, $z_\beta$ that displace the system away
from the collinear configuration parallel to a manifold of cgs. $H$
contains no terms of the form $z_\alpha^2$, $z_\beta^2$, but only
term $z_\alpha^2 z_\beta^2$, higher order terms and terms containing
other displacements. Let us displace the system in the $\beta$
direction to say $z_{0\beta}$. Then $z_\alpha$ has to stay zero for
the system to remain in cgs. But around the new point, the mode
along the $\alpha$ direction would become $z_{0\beta}^2 z_\alpha^2$
and so $\lambda \sim z_{0\beta}^2 \sim \theta^2$. There are of
course all the other modes that need to be taken into account but it
is reasonable that this result will not change. To further check
this, we expanded the potential around one particular noncollinear
cgs, one in which each primitive unit cell has the same spin
configuration, with the
spins displaced by an angle $\theta$ from the collinear
configuration. The results are shown in Figure \ref{fig_Bands_all}a
and \ref{fig_Bands_all}b (with $\theta=0.3$). For given $n$ and
$\theta$ the ``spin-wave'' bands are the union of b) with $(n-2)$
copies of a). We see that two zero bands, which together contain N
modes, become nonzero (except for special directions that are of
measure zero). We further find that stiffnesses of these soft modes
go as $\lambda \sim \theta^2$.

\subsection{Quantum ground state}
Here we show that the ``power-counting" variational argument used
for the single tetrahedron goes through for the full lattice with
only slight modifications. We find that the ground state
wavefunction is localized around a collinear configuration for small
$g$ and find its spread as a function of $g$ and $n$.

We consider a variational wavefunction that is localized near
$\theta=0$, spreading in angle by $\gamma$ in each of the $\sim nN$
directions available to it.  The kinetic energy of such a state is
$\sim gnN/\gamma^2$.  The typical angle $\theta$ in this state is
the (Pythagorean) sum of $(n-1)$ angles of displacement that are
mutually perpendicular to each other and each of order $\gamma$, so
$\theta\sim\gamma\sqrt{n-1}$. The kinetic energy is $\sim
Nn^2g/\theta^2$. The are $N$ soft modes so the potential energy is
$\sim N\theta \sqrt{g}$. Minimizing the total energy with respect to
$\theta$ gives the same estimate as before for its typical value in
the ground state:
\begin{equation}
\theta_0 \sim n^{\frac{2}{3}}g^{\frac{1}{6}}~.
\end{equation}
Thus we see that the ground state is collinear ($\theta_0\rightarrow
0$) in the $g\rightarrow 0$ limit.  The crossover from collinear
ordering (small $\theta_0$) to a strongly disordered state can be
defined as occurring at $\theta_0\sim 1$, which puts the crossover
in the ground state at
\begin{equation}
g_c\sim n^{-4}
\end{equation}
for large $n$.

\subsection{T$>$0}
We would like to extend the analysis done for the tetrahedron. What
is more complicated here, is that the soft modes are not all the
same and further they depend in some complicated way on the
displacement. We found that for one particular displacement, their
stiffnesses go as $\lambda_i = k_i \theta^2$ where $k_i$'s are shown
on Figure \ref{fig_Bands_all}c. To proceed we are going to make the
following assumption: Given some general displacement that is
characterized some mean $\theta_{rms}$ of displacements of all
spins, assume that this distribution of $k_i$'s is not going to
change very much, at least in a qualitative way.

The probability of finding the system in some cgs, treating the
perpendicular directions in harmonic approximation is given by
\[ P \sim \prod_i \frac{1}{\sinh(\sqrt{\lambda_i} \frac{\sqrt{g}}{T})} \]
where the $\lambda_i$'s are the eigenvalues of the matrix of the
second derivatives of the potential in the perpendicular directions.
With our approximations the probability of finding the system with
some $\theta = \theta_{rms}$ is
\[ P(\theta) \sim \prod_i \frac{\theta^{N(n-2)}}{\sinh(\sqrt{k_i}\theta\frac{\sqrt{g}}{T})} \]

To find the localization, we need to find the behavior of this
function. We find that for large $n$ it is sharply peaked at
\[\theta_f = n \frac{T}{\sqrt{g}}\] To see this differentiate its
logarithm with respect to $\theta$
\begin{eqnarray*}
\frac{d}{d\theta} \log P(\theta) = \frac{1}{\theta} \left( N(n-2) -
\sum_i \frac{\sqrt{k_i}\theta\frac{\sqrt{g}}{T}}{\tanh \left(
\sqrt{k_i}\theta\frac{\sqrt{g}}{T}\right)} \right)
\end{eqnarray*}
and analyze the resulting function for large $n$. The $k$'s satisfy
$0<k_i<1$. For $\theta\frac{\sqrt{g}}{T} \ll 1$ the second term is
approximately $N$ and so the derivative is positive and the function
is increasing. For $\theta\frac{\sqrt{g}}{T} \gg 1$ for most of the
values of $i$, the term under summation sign is approximately
$\sqrt{k_i} \theta\frac{\sqrt{g}}{T}$ and so the second term is
approximately $N \theta\frac{\sqrt{g}}{T}$. It is a bit smaller
because $k_i < 1$ but of that order. The derivative is zero when
$\theta\frac{\sqrt{g}}{T} \sim n$ and is negative for larger values
of $\theta$ and so as said, the function is peaked at $\theta_f$.
Evaluating the second derivative we find that the function is
sharply peaked with ratio of the width to $\theta_f$ being
$\sqrt{\frac{1}{Nn}}$. Therefore the system is localized up to
approximately $\theta_f$ just as in the case of the single
tetrahedron.

As $n$ decreases the $P$ becomes more and more distributed around
zero and at $n=3$ the it is only a decreasing function. Nevertheless
most of it's weight is still in the region approximately $<
\theta_f$ and so the system is still localized approximately up to
$\theta_f$.

\subsection{Scaling argument}
The previous argument shows that the spins localize, but assumes
that the distribution of $k_i$ doesn't change significantly for
other displacements. In this section we relax this condition and
only assume that $\lambda \sim \theta^2$. We will show that if the
system localizes it depends on $g$ and $T$ only through $\theta_0 =
T/\sqrt{g}$.

We want to know the probability that the $\theta$'s have rms
$\theta_{rms} = \theta_r$. This is given by
\begin{eqnarray*}
&&P(\theta_r, \theta_0) = \\
&& \frac{ \int \prod d\theta (\prod_i \sinh (\sqrt{\lambda_i
(\lbrace \theta \rbrace)} / \theta_0))^{-1} \delta (\sqrt{\sum
\theta_i^2} - \theta_r) } { \int \prod d\theta (\prod_i \sinh
(\sqrt{\lambda_i (\lbrace \theta \rbrace)} / \theta_0))^{-1} }
\end{eqnarray*}

The $\int \prod d\theta$ is symbolic, it means to sum over all
ground state configurations, in the neighborhood of the collinear
configuration. By changing variables and using the scaling
property of the eigenvalues it is easy to see that
\[ P(c \theta_r, \theta_0) = P(\theta_r, c \theta_0) \]
Thus the $P$ and depends only on $\theta/\theta_0$ (and $n$).
Therefore the localization depends on $g$, $T$ only through
$\theta_0$.

\subsection{Spin wave theory}
We now consider a, quadratic, spin wave analysis about the collinear
cgs to see if a state selection in this approximation occurs. We
note that this should be asymptotically accurate at small $g$ and
that in the spin system this corresponds to the $1/S$ computation
which {\it does} select a subset of the collinear states
\cite{Henley}.

A given collinear cgs can be concisely specified, upto a global rotation,
by writing the spins as $\vec{S}_i = \eta_i \hat{z}$ where
$\eta_i = \pm 1$. Fluctuations about such a configuration can
be parametrized as $\vec{S}_i = (\vec{x}_i, \eta_i \sqrt{1-x_i^2})$
where $\vec{x}_i = (x_i^1,\ldots,x_i^{n-1})$. Expanding the square root
to second order in $x$, the nearest neighbor interaction can be written
as
\[ V = \sum_{<i,j>} \eta_i \eta_j + \sum_{<i,j>} \vec{x}_i \cdot
\vec{x}_j + \sum_i \eta_i x_i^2 \sum_{j \mbox{ } nn \mbox{ } i}
\eta_j \ .
\]
In every collinear cgs the first term is the same and the sum in the last
term is $\sum_{j \mbox{ } nn \mbox{ } i} \eta_j = -2\eta_i$ which
eliminates all $\eta$s from the last term. Thus the potential expanded to
the second order around a given collinear configuration is
completely independent of the choice of collinear configuration. Hence any
selection beyond collinear ordering must come from higher orders in the
expansion and thus is a weaker effect than one might have guessed {\it a
priori}.

\section{Numerical Simulations}
In this section we use imaginary-time path-integral quantum Monte
Carlo simulations to study the ordering of our quantum rotors on the
pyrochlore lattice in the Heisenberg case $n=3$.  This method is
based on writing the partition function as the trace over all states
and inserting $N_t-1$ additional resolutions of the identity to
obtain $N_t$ copies (``time slices'') of the pyrochlore lattice
model, coupled ferromagnetically along the imaginary time direction.
The case of one time slice is the classical model that does not
order. We find that an ordered phase does appear already in the case
of two time slices, although it is restricted to very low
temperature. As the number of time slices is increased, the ordered
phase apparently becomes more stable.

In principle, to get the quantitatively correct behavior of our
quantum Hamiltonian, we should take the continuum (Hamiltonian)
limit of $N_t\rightarrow\infty$ (with the appropriate scalings of
$K_0$ and $K_1$). However, doing this extrapolation properly is a
computationally demanding task that we have not seriously attempted.
Instead we have simulated primarily the case $N_t=8$, mapping out a
portion of its phase diagram and characterizing the phase transition
in to the spin-nematic ordered phase. We find that the ordering
transition is first-order, and that the phase diagram is indeed
qualitatively similar to the crossover diagram we obtained for a
single tetrahedron.  We expect that these conclusions are correct
also in the Hamiltonian limit.

\subsection{Quantum Monte Carlo}
The system we simulate has partition function
\begin{eqnarray}
 Z &=& \int \prod_{k,i} d\vec n_{k,i} e^{S} \\
 S &=& K_0 \sum_{k,i} \vec n_{k,i}\cdot\vec n_{k+1,i} - K_1 \sum_{k,<i,j>} \vec n_{k,i}\cdot\vec n_{k,j} \\
 K_0 &=& \frac{N_t T}{g} \label{eq_K0} \\
 K_1 &=& \frac{1}{N_t T} \label{eq_K1}~,
\end{eqnarray}
where $\vec n_{k,i}$ is a unit vector at site $i$ in time slice $k$.
The nearest neighbor pairs on the pyrochlore lattice are denoted
$<i,j>$, and the relations between the couplings here and $T$ and
$g$ are shown.  The latter are valid in the large $N_t$ limit,
although we will use them as an approximation for finite $N_t$.  The
samples are of size $N_x^3$ primitive unit cells (thus $4N_x^3$
spins) in each time slice, with periodic boundary conditions.

\subsection{Order parameter}
The spin-nematic order parameter that we use is the symmetric
traceless tensor
\[ Q^{\alpha \beta} = \frac{1}{4N_tN_x^3} \sum_{k,i} (n_{k,i}^\alpha n_{k,i}^\beta - \frac{1}{3} \delta^{\alpha \beta})~,\]
where $\alpha$ and $\beta$ run over the three directions in spin
space.  Note this order parameter is defined as a sum over all of
space and imaginary time at one instant during our simulation.  To
make a dimensionless combination that is sensitive to the ordering,
we use the generalized ``Binder ratio''
\[ q_3 = \sqrt{6}\langle\frac{Tr Q^3}{(Tr Q^2)^\frac{3}{2}}\rangle~,\]
where the average here is over different instantaneous measurements
of $Q$ during the simulation.  This quantity is a measure of the
degree of collinear spin-nematic order.  It is zero for randomly
oriented spins and increases as one approaches and enters the
ordered phase, taking the value one in the well-ordered limit.  It
is thus a good quantity for doing a finite-size scaling analysis of
the phase transition, as we show below.  Note that the average value
of the third power $Tr Q^3$ of the order parameter does not vanish
in this model.  This reflects the fact that its Landau theory has a
cubic term and thus the phase transition is expected to be of first
order, just as in the case of the isotropic-to-nematic transition in
3D liquid crystals.

\subsection{Phase diagram and the order of the transition}
We first establish the existence of the ordered phase. We simulate
the system with sizes $N_t=8$, $N_x=4,5,6,7$ for $K_0=3$ and
$K_1=10.1$ or $K_1=10.2$. In the process, for every configuration in
the run, we save the ``action'' $S_0$ due to the couplings along the
imaginary
time direction and $S_1$ due to the space direction. 
This allows us to obtain estimates of $q_3$ for not only the values
of the couplings simulated, but also for nearby values of the
couplings by each spin configurations its Boltzmann weight $e^{K_0
S_0 + K_1 S_1}$ in calculating the averages of $q_3$.

The results for $q_3$ for $K_0=3$ and a range of $K_1$ near the
phase transition are shown in Fig. 5\label{fig_q3s}.  A crossing of
the curves for the different sizes is clearly seen at the estimated
``critical'' value of $K_{1c}\cong 10.20$.

\begin{figure}[h]
\epsfxsize=4in \centerline{\epsfbox{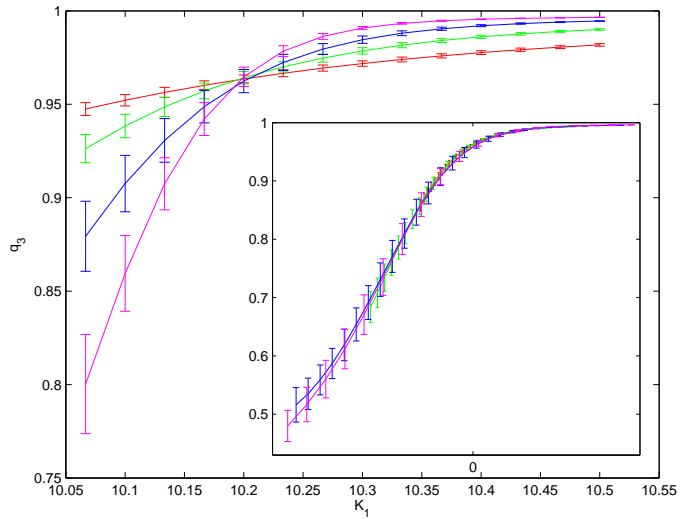}} \caption{The
values of $q_3$ for sizes $N_x=$4,5,6,7 (as $N_x$ increases, the
curves become steeper).  The curves cross at one point, which is the
estimated location of the phase transition.  The inset shows $q_3$
plotted vs. the scaling variable $(K_1-K_{1c})N_x^3$ appropriate for
a first-order transition. } \label{fig_q3s}
\end{figure}

Next we find how the order parameter scales with the size of the
system. We rescale $(K_1-K_{1c}) \to (K_1-K_{1c}) N_x^{1/\nu}$ and
tune $\nu$ to see when the curves of $q_3$ for different sizes
align. We find that they align well for $\nu=1/3$, as shown in the
inset to Figure 5. Thus we can write $q_3 = f((K_1-K_{1c}) N_x^d)$
where $d=3$ is the dimension of the system. This is the scaling
expected for a first order phase transition.

\begin{figure}[h]
\epsfxsize=4in \centerline{\epsfbox{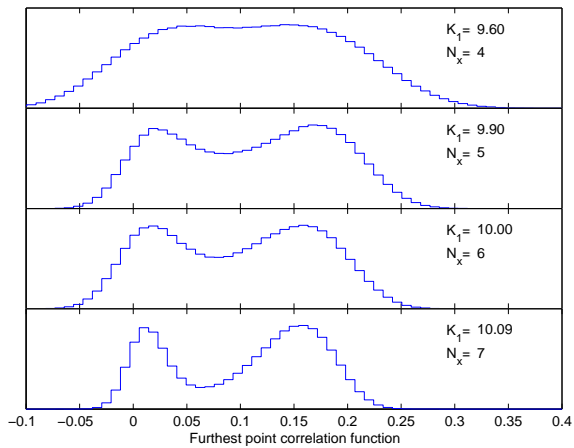}}
\caption{Probability distributions of the furthest-point correlation
function for $N_t=8$, $K_0=3$ at the values of $K_1$ where the two
peak heights match. The peaks become sharper and the minimum in
between deeper with increasing system size, as expected for a
first-order transition.} \label{fig_HistSamePeaks}
\end{figure}

To give further evidence of for the first order phase transition we
calculate the furthest point correlation function defined as follows
\[\frac{3}{2}<(\vec{n}_{t,x,y,z}\cdot\vec{n}_{t+N_t/2,x+N_x/2,y+N_y/2,z+N_z/2})^2>-\frac{1}{2}\]
where the average is over all spins at one instant during the
simulation.  If the transition were first order, the finite size
system at the transition point would jump back and forth between the
ordered and disordered phase. Thus if we plot the histogram of these
correlations we should see two peaks. These should get sharper as we
increase the size of the system. The numerical results are shown in
Figure \ref{fig_HistSamePeaks} where we indeed see two peaks, which
are getting sharper with increasing size. These results, combined
with our expectations from the Landau theory and the finite-size
scaling of $q_3$ give strong evidence that the transition is first
order. The place where we are showing this corresponds to a $g$ near
the maximum of $T_c$, but we expect this first order character will
also remain elsewhere on the phase boundary, including at the $T=0$
quantum phase transition at $g_c$.

\begin{figure}[h]
\epsfxsize=4in \centerline{\epsfbox{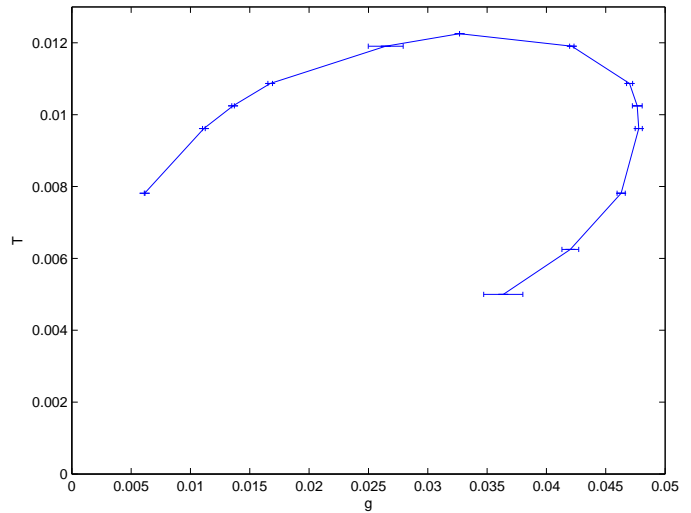}}
\caption{The estimated phase transition curve for the $N_t=8$,
$N_x=5$ system. The strong reentrance at low $T$ on high $g$ side of
the phase diagram is likely an artifact of the very coarse imaginary
time slicing that is done there with these parameters. }
\label{fig_TransitionCurve}
\end{figure}

Now that we have explored one point in the phase diagram, we would
like to look at more of them and find the shape of the phase
transition curve. In order to simplify the calculations we will look
only for one size $N_x=5$, $N_t=8$ and say that the point is a point
of phase transition if its $q_3$ value is the same as that at the
crossing point in Fig. 5. Finding a point sufficiently close to the
phase transition curve we reweight configurations as described above
to find a point with this value of $q_3$. The results are shown in
Figure \ref{fig_TransitionCurve}. The range that is readily
accessible to our numerical simulations corresponds to the higher
temperature part of the phase boundary.  We see that it has
qualitatively the same shape as that of our single tetrahedron
crossover diagram, and in particular it shows the reentrant behavior
on varying $g$ at a fixed $T$ (in practice this is varying $K_0$ at
a fixed $K_1$).  The strong reentrance also seen at low $T$ on the
high $g$ side of the phase diagram is likely an artifact of the very
coarse imaginary time slicing that is done there with these
parameters. We note that the transition temperature is very low, of
order $T_c \sim 1\%$ of $J$ ($J=1$ in this paper) and is a
manifestation of strong frustration in this system.

\section{Large n analysis}
In this section we analyze the system in the limit $n \to \infty$ where
rotor models are known to be solvable by a saddle point method.
In the previous analysis we found that in this limit the spins don't
order at any $g$ and we will recover this result. Additionally,
we obtain the spin correlations in the disordered phase.

A brief recap of the analysis: We write the partition function as a
path integral and impose the fixed length constraint by introducing a
Lagrange multiplier field. The resulting action is quadratic in spins
which can then be integrated out to get an effective action for the
Lagrange multiplier field with an overall coefficient of order $n$
which thus enables a saddle point treatment.
At $T=0$ the saddle point condition takes the form
\[ \frac{1}{2N}\sum_{q,\sigma}
\frac{\sqrt{g}}{\sqrt{\mu_{q}^{\sigma} + \lambda}} = 1 \] where
$\lambda$ is the uniform, saddle point value of the Lagrange
multiplier field, $\sigma$ runs over the four bands of the
pyrochlore lattice and $\mu_q^{\sigma}$ are the eigenvalues of the
adjacency (interaction) matrix for a lattice of $N$ sites. As the
the lowest two bands are flat, with $\mu^\sigma$ independent of
${\bf q}$, this equation has a solution with $\lambda >0$ for any
nonzero value of $g$. Hence the system is always quantum disordered
at $n = \infty$.

We will not write the resulting correlations of the spins explicitly.
Instead it is instructive to write the correlations of the field that
is obtained from them as follows. Let $\vec{n}$, defined at any site, be a
vector pointing from one tetrahedron to the other (this can be
chosen consistently as the tetrahedra surround sites of the bipartite
diamond lattice).
Then define the vector field $\vec{B}^a(x) = \vec{n}(x) S^{a}(x)$. The
significance of this field lies in the fact that it has zero lattice,
and thus coarse-grained, divergence in the classical ground state
manifold. As discussed in Ref.~\cite{isakov1} this field exhibits
dipolar correlations in the $T \rightarrow 0$ limit in the classical
problem. In our quantum problem,  at small $g$ and small $q$, we find
that it exhibits the ground state correlations
\begin{eqnarray*}
&&\langle B_i^a(q) B_j^b(-q) \rangle = \\
&&\delta^{ab} \left( \frac{\delta_{ij}q^2 - q_i
q_j}{q^2}\frac{g}{\omega^2 + g^2} + \frac{q_i q_j}{q^2}
\frac{g}{\omega^2 + g^2 + g q^2}\right)
\end{eqnarray*}
where we have ignored, for simplicity, numerical factors that appear in
front of $g$.

These forms reduce at equal times and $g=0$ to the dipolar forms
derived in \cite{isakov1}. At nonzero $g$ we see that the
correlations decay exponentially in space with a correlations length
of order $g^{-1/2}$ and exponentially in time with a gap of order
$g$. Observe that the $n=\infty$ problem does not contain any trace
of the nematic ordering that exists at finite $n$. Individual terms
in the $1/n$ expansion for the correlation function can be seen to
be well behaved \cite{ghs} and so it will take an analysis of the
series to reproduce the instability that we have obtained earlier.
This is a challenge for future work.

\section{Conclusions}

We have fully explored the local ordering of a single tetrahedral
unit of four neighboring rotors.  They order collinearly, with two
rotors pointing along one direction and the other two in the
opposite direction. Thus an {\it axis} of ordering is chosen, and
since in the full pyrochlore lattice this tetrahedron shares one
rotor with each of its neighbor tetrahedra, this axis of ordering is
uniform throughout the lattice in the ordered phase, which thus has
spin-nematic, or collinear order.  We have demonstrated this
spin-nematic ordering within a quantum Monte Carlo simulation. There
might also be long-range order in which rotors point in which
direction along this axis, and/or in which pairs of rotors are
parallel or antiparallel. We have not yet been able to determine
whether our ordered phase has any of these latter types of
sublattice order in addition to its collinear order. We find that
the collinear order shows up quite robustly in our quantum Monte
Carlo simulations, while any other long-range correlations that
might be present appear to be much weaker and difficult to detect,
if they are indeed there.

We note that the considerations outlined here apply straightforwardly
to the the two-dimensional checkerboard lattice which is the planar
analog of the pyrochlore lattice. By contrast an analysis of the
rotor models on the kagome lattice will require a fresh analysis.

We thank the NSF for support through MRSEC grant DMR-0213706.


\begin{thebibliography}{10}
\bibitem{fn-xy} The pattern of the ordering is not known.
\end{thebibliography}
\end{document}